# Butterfly-shaped magnetoresistance in triangular-lattice antiferromagnet $Ag_2CrO_2$

**Hiroki Taniguchi[1], Mori Watanabe[1], Masashi Tokuda[1], Shota Suzuki[1], Eria Imada[1], Takashi Ibe[1], Tomonori Arakawa[1,2], Hiroyuki Yoshida[3], Hiroaki Ishizuka[4], Kensuke Kobayashi[1,5,6], and Yasuhiro Niimi[1,2,*]**

[1]Department of Physics, Graduate School of Science, Osaka University, Toyonaka 560-0043, Japan.

[2]Center for Spin Research Network, Osaka University, Toyonaka 560-8531, Japan.

[3]Department of Physics, Graduate School of Science, Hokkaido University, Sapporo 060-0810, Japan

[4]Department of Applied Physics, Graduate School of Engineering, The University of Tokyo, Bunkyo, Tokyo 113-8656, Japan

[5]Department of Physics, Graduate School of Science, The University of Tokyo, Tokyo 113-0033, Japan.

[6]Institute for Physics of Intelligence, Graduate School of Science, The University of Tokyo, Tokyo 113-0033, Japan.

*To whom correspondence should be addressed. E-mail: niimi@phys.sci.osaka-u.ac.jp (Y. N.)

Spintronic devices using antiferromagnets (AFMs) are promising candidates for future applications. Recently, many interesting physical properties have been reported with AFM-based devices. Here we report a butterfly-shaped magnetoresistance (MR) in a micrometer-sized triangular-lattice antiferromagnet $Ag_2CrO_2$. The material consists of two-dimensional triangular-lattice $CrO_2$ layers with antiferromagnetically coupled $S = 3/2$ spins and $Ag_2$ layers with high electrical conductivity. The butterfly-shaped MR appears only when the magnetic field is applied perpendicularly to the $CrO_2$ plane with the maximum MR ratio ($\approx$ 15%) at the magnetic ordering temperature. These features are distinct from those observed in conventional magnetic materials. We propose a theoretical model where fluctuations of partially disordered spins with the Ising anisotropy play an essential role in the butterfly-shaped MR in $Ag_2CrO_2$.

**Introduction**

Compared to ferromagnets (FMs), antiferromagnets (AFMs) are much more difficult to manipulate the magnetic state. Thus, it has been believed that AFMs do not seem to have any applications[1]. In the field of spintronics, since 1990s, AFMs have been mainly used as a pinned layer in spin-valve devices where parallel and antiparallel magnetic states of two FM layers induce a large resistance change and one of them has an AFM layer underneath in order to raise the magnetic coercivity[2,3]. Recently, AFMs have played a more essential role in spintronics[4-6] because AFM-based spintronic devices are not only robust for external magnetic fields[7] but also generate many interesting phenomena such as very fast domain wall motions[8,9], large anomalous[10] and spin Hall effects[11], antiferromagnetic dynamics in the THz range[12,13], and so on.

When antiferromagnetically-coupled spins are arranged in a triangular shape, some interesting physical properties can be expected. For example, $Mn_3X$ (X = Sn, Ge) has an AFM ordering phase with 120° structure above room temperature and the remnant magnetic moment in the AFM state is very small ($\approx 0.002\mu_B$ per Mn atom where $\mu_B$ is the Bohr magneton)[14-17]. Nevertheless, it shows an extremely large anomalous Hall effect, which seems to be inconsistent with the commonly believed picture: the anomalous Hall effect is proportional to the total magnetization[18]. A fictitious field due to the Berry curvature is in charge of the anomalous Hall effect[14-17]. In addition, a peculiar spin Hall effect, namely magnetic spin Hall effect, has also been reported in this material[19,20]. For a better understanding of the above peculiar phenomena, it is important to further perform magnetic transport measurements in triangular lattice AFM systems with electrical conductivity.

Ag$_2$CrO$_2$ is a *highly* conductive triangular lattice AFM with the transition temperature $T_N$ of 24 K[21]. The crystal structure of Ag$_2$CrO$_2$ is shown in Fig. 1(a). The Cr site has an $S = 3/2$ moment, which is arranged in a triangular shape. It is known that the coupling between the two Cr sites in the two-dimensional (2D) plane is antiferromagnetic[21]. The conductive Ag$_2$ layers which are coupled to the magnetic CrO$_2$ layers enable us to investigate this spin system with electrical transport measurements. Such an electronic property is unique and advantageous compared to most of triangular lattice AFMs which are insulators or have very low electric conductivity.

According to the neutron[22] and muon spin resonance ($\mu$SR) experiments[23], Ag$_2$CrO$_2$ has a unique thermodynamic property, i.e., partially disordered (PD) state with 5 sublattices, at finite temperatures[24,25] [see Fig. 1(c)], which is different from the above 120° structure. In principle, the PD spin acts as a free spin since all the interactions from the nearest, the second nearest, and even the third nearest neighbors are canceled out. Thus, there should be negligibly small magnetization below $T_N$. Nevertheless, this compound has a finite magnetization ($\approx 0.08\mu_B$ per Cr atom) at zero field[21,24]. The origin of the small magnetization is still unclear. This system is also quite interesting from the perspective of spin fluctuations in the PD state.

In this work, we performed magnetotransport measurements in order to investigate the impact of the PD spin fluctuations on the electrical transport property using a micrometer-sized Ag$_2$CrO$_2$; it is close to the single crystal. We find a butterfly-shaped magnetoresistance (MR) when the magnetic field is applied along the *c*-axis. Unlike the case of conventional magnetic systems, the amplitude of the butterfly-shaped MR is small at low temperatures and takes a maximum at around $T_N$, and disappears above $T^*$ = 32 K. The result coincides with the MR by magnetic fluctuation, which is based on the 2D magnetic system with the uniaxial anisotropy.

**Sample Fabrications and Experimental Setup**

Polycrystalline $Ag_2CrO_2$ samples were obtained by encapsulating a mixture of Ag, $Ag_2O$, and $Cr_2O_3$ powders in a gold cell, and by baking them at 1200°C for 1 hour under a pressure of 6 GPa[21]. The polycrystalline samples were then pounded on a glass plate in order to obtain small pieces of $Ag_2CrO_2$. The small grains were picked up with a scotch tape and pasted onto a thermally oxidized silicon ($SiO_2$/Si) substrate with several 100 nm thick gold marks. After removing the scotch tape from the substrate, another $SiO_2$/Si substrate without any gold marks was pushed onto the substrate with the $Ag_2CrO_2$ flakes and the 100 nm thick gold marks. In this process, $Ag_2CrO_2$ flakes thinner than ≈100 nm are left on the $SiO_2$/Si substrate with the gold mark, and relatively thick $Ag_2CrO_2$ flakes are transferred to the substrate without the gold marks[25]. As a result, $Ag_2CrO_2$ flakes with a few micrometer-size remain on the substrate with the gold marks. We have fabricated 10 different devices and always observed the butterfly-shaped structure as detailed in the next section. The thicknesses of these flakes (80~120 nm) were confirmed by a commercially available atomic force microscope.

To confirm if the $Ag_2CrO_2$ flakes used for transport measurements are close to a single crystal, we took scanning transmission electron microscope (STEM) images shown in Figs. 2(a) and 2(b). As confirmed with the X-ray diffraction pattern for the polycrystalline samples, there is a single $Ag_2CrO_2$ phase[21]; no other phases such as $AgCrO_2$[26], which is a ferroelectric material. The $Ag_2$ and Cr layers [see Fig. 1(a)] are alternatively stacked perpendicularly to the $SiO_2$/Si substrate. On top and bottom of the $Ag_2CrO_2$ flake, some Ag clusters segregated can be seen. However, inside the $Ag_2CrO_2$ flake, there is no obvious cluster, indicating that the flake is close to a single crystal.

To perform the transport measurement, we deposited 150 nm thick Cu electrodes to the micrometer-sized $Ag_2CrO_2$ films, using the standard electron beam lithography and a Joule heating evaporator[25]. A typical device is shown in the inset of Fig. 1(b). The lateral size and the thickness of the $Ag_2CrO_2$ flake are a few μm and 100 nm, respectively. The contact resistance between $Ag_2CrO_2$ and Cu is less than 1 Ω, which is comparable to a contact resistance for normal metallic junctions with almost the same junction area. This is also consistent with the STEM image in Fig. 2(c): the segregated Ag part on the left top side is continuously connected to the most top part of the $Ag_2CrO_2$ flake, showing that the Ag layer is exposed after the fabrication. The transport measurements have been carried out using an ac lock-in amplifier and a $^4$He flow cryostat.

**Experimental Results**

In Fig. 1(b), we plot the longitudinal resistivity $\rho_{xx}$ of the $Ag_2CrO_2$ device as a function of temperature. The overall temperature dependence is metallic and there is a large resistivity drop at around $T_N$ determined from the heat capacity measurements for polycrystalline $Ag_2CrO_2$ samples[21]. This resistivity change can be explained by spin fluctuation: fluctuations of paramagnetic spins at the Cr sites above $T_N$ are strongly suppressed below $T_N$ and the scattering rate of the conduction electrons is reduced. We also note that $\rho_{xx}$ at 5 K obtained for the thin film device is about 10 times smaller than that for the polycrystalline bulk samples[21,25]. From this result, we can argue that the thin film devices obtained with the mechanical exfoliation technique include much less grain boundaries compared to the polycrystalline bulk samples, resulting in a much better quality of $Ag_2CrO_2$. This is also supported by the STEM image in Fig. 2.

For the $Ag_2CrO_2$ device, we performed MR measurements with three different magnetic field ($B$) directions, i.e., $x$ (in-plane along the current direction), $y$ (in-plane perpendicular to the current direction), and $z$ (out-of-plane along the $c$-axis, i.e., $z \parallel c$) directions. Figures 3(a)-(h) shows the MRs, $\frac{\rho_{xx}(B)-\rho_{xx}(0)}{\rho_{xx}(0)}$, along the three directions measured at several different temperatures. At $T$ = 5 K ($\ll T_N$), a clear positive MR is observed at high magnetic fields when $B$ is applied along the $z$-direction. With increasing temperature, the positive slope becomes flatter. At 25 K ($\approx T_N$), the MR shows the negative sign. This trend is explained by the competition of two different mechanisms for MR, i.e., the ordinary MR and the MR related to spin fluctuation. The positive MR at $T \ll T_N$ is related to the ordinary MR by the Lorenz force because the magnetic fluctuation is suppressed in this temperature region. On the other hand, when $T \sim T_N$, the magnetic scattering by thermal fluctuation of spins is enhanced. This contribution to the resistivity is suppressed by the magnetic field perpendicular to the plane, producing the negative MR.

Another unique feature is the butterfly-shaped MR at $B \approx \pm 0.5$ T. The amplitude of the butterfly-shaped MR is small when $T \ll T_N$. As we approach $T_N$, it becomes larger and takes a maximum at 25 K ($\approx T_N$). The maximum value reaches more than 10% at $B = 0.5$ T, which is unusually large for conventional ferromagnetic materials[27,28]. As we raise the temperature further, the amplitude of the MR suddenly decreases and becomes zero above $T^* = 32$ K.

In contrast to the MR along the $z$-direction, such a drastic temperature dependence of MR has not been observed when $B \parallel x$ and $B \parallel y$, although a small negative MR can be seen below $T_N$. The $B$-angle dependence of the MR, which has never been studied for polycrystalline bulk $Ag_2CrO_2$[21,24].

To evaluate the butterfly-shaped MR observed only for $B \parallel z$, we define the amplitude of the buttery-shaped MR, i.e., $\Gamma \equiv \frac{\rho_{xx}^{\text{upper}}(B_c) - \rho_{xx}^{\text{lower}}(B_c)}{\rho_{xx}(0)}$, and the corresponding magnetic field ($B_c$), as illustrated in the inset of Fig. 4(a). $\Gamma$ has a small value at low temperatures and takes a maximum (15%) at around $T_N$. It still has a finite value even above $T_N$ and finally disappears at $T^*$. $B_c$ in Fig. 4(b) is almost constant up to $T \approx 22$ K, and starts to decrease with increasing temperature and disappears at $T^*$.

Similar MR effects are often observed not only in current-in-plane (CIP) giant-magnetoresistance (GMR) devices[27,28] but also in ferromagnetic[29,30] and even antiferromagnetic materials[31]. However, the present butterfly-shaped MR is essentially different from them. While commoly-used CIP-GMR devices have in-plane magnetization whose direction is the same as the current direction, $Ag_2CrO_2$ has a penpendicular magnetization to the basal plane and the current direction. An MR in conventional magnetic materials depends on the relative angle of magnetic domains, which is tuned by *B*. The amplitude of the MR is at most less than 1%[30] at *B* = 0.5 T. It decreases with increasing temperature and becomes zero above the transition temperature. In the butterfly-shaped MR, however, $\Gamma$ has a maximum value of 15% near the transition temperature, which cannot be expected in conventional magnetic materials[29-31]. These experimental facts indicate that spin fluctuations of the PD state are strongly related to the butterfly-shaped MR. To our knowledge, this is the largest MR value induced by spin fluctuations.

**Discussions**

What is the origin of the butterfly-shaped MR? As mentioned in the introduction, $Ag_2CrO_2$ is basically an antiferromagnetic metal and the PD spin behaves as a free spin. Thus there should be no spontaneous magnetization, but it has been established that the polycrystalline $Ag_2CrO_2$ has a small but finite spontaneous magnetization below $T_N$[21-24]. The origin of the spontaneous magnetization is still an unsolved problem, but the PD spins should play an essential role in the butterfly-shaped MR. Here we assume the same magnetic state for the thin flim device as the bulk sample, although the magnetization measurement has not been performed. We also recall that the butterfly-shaped MR in the present work appears only when *B* || *z*. These features imply that the uniform magnetic moment is along the *z*-axis and has a strong uniaxial anisotropy. As illustrated in Fig. 5, we assume that the PD spin is canted and has a small magnetic moment along the *c*-axis. In addition, we also assume that the magnetic fluctuation of the PD spin is relatively large, since the PD spin flips only at 0.5 T. In such a situation, *B* suppresses the spin fluctuations when it is parallel to the moment direction, while $B$ causes a spin flip when it is antiparallel to the moment direction. Thus, a negative and butterfly-shaped MR can be explained by the suppression of spin fluctuation and the spin-flip process induced by $B$, respectively.

As one of the possible models to explain the butterfly-shaped MR, we consider a 2D ferromagnetic spin system with the Ising anisotropy. Under a small magnetic field ≤1 T, we assume that the ferromagnetic magnon well approximates the low-energy magnon states of $Ag_2CrO_2$. The Hamiltonian is given by

$$H = -J\sum_{\langle i,j \rangle} \mathbf{S}_i \cdot \mathbf{S}_j - \Delta \sum_i S_i^z S_i^z - B \sum_i S_i^z \quad (1)$$

where $J$ (> 0) is the ferromagnetic exchange coupling between the nearest neighbor sites, $\Delta$ (> 0) is the Ising anisotropy, and $i$ and $j$ are the site numbers. $B$ is applied perpendicular to the 2D plane (i.e., $B \parallel z$). The uniaxial anisotropy lifts the Goldstone mode, producing a spin gap proportional to the anisotropy energy $2\Delta$ ($\propto |B_c|$), as illustrated in Fig. 6. For the positive magnetization, the spin gap increases (decreases) by applying the positive (negative) magnetic field and becomes zero when $B = B_c$ (< 0). The suppression of the magnetic fluctuation for $B > 0$ and the spin flip at $B = B_c$ (< 0) are intuitively explained by the spin gap modulated by $B$. The exactly same scenario is valid for the negative magnetization just by inverting the sign of $B$.

To see how $B$ suppresses the MR, we compute the elastic scattering rate $\frac{1}{\tau_{\text{mag}}} \propto \rho_{xx}$ due to spin fluctuations at finite $T$ by Born approximation[32]:

$$\frac{1}{\tau_{\text{mag}}} \propto T \left\{ F_1 \left(1 + \frac{2\Delta + \mu_{\text{eff}}B}{2Jk_F^2 a_0^2}\right) - F_2 \left(1 + \frac{2\Delta + \mu_{\text{eff}}B}{2Jk_F^2 a_0^2}\right) \right\} \quad (2)$$

where $k_F$ is the Fermi wave number, $\mu_{\text{eff}}$ is the effective ferromagnetic moment, $a_0$ is the lattice constant between the neighboring effective ferromagnetic moments, $F_1(x) = \frac{1}{\sqrt{x^2-1}}$ and $F_2(x) = \frac{x}{\sqrt{x^2-1}} - 1$. A fit to the experimental data in Eq. (2) is shown in Fig. 3(e); the butterfly-shaped MR near $T_N$ is well-reproduced by our theoretical model. We have roughly estimated both $J$ and $\Delta$ to be about 10 K, which is reasonable for the present case[32].

Finally, let us metion the relation between $T_N$ and $T^*$. In the present experiment, the butterfly-shaped MR takes a maximum at around $T_N$ and vanishes at $T^* = 32$ K. Originally, $T_N$ is determined from the peak position of the heat capacity measurement as shown in Ref. 21. A long range ordering emerges below $T_N$ (~24 K), but some short range ordering with magnetic fluctuations may grow even above $T_N$. This was already pointed out by Sugiyama *et al*. from zero-field $\mu^+$SR measurements[23]. They argued that the phase with the PD spins already grows above $T_N$ and vanishes $T^* = 28$ K. Those tendencies are indeed consistent with our experimental data. In addition, it is also known that $T_N$ is slightly shifted to the higher temperature side, by applying the magnetic field[21]. Thus, it seems to be reasonable that $T^*$ determined from the present MR measurement is higher than $T^*$ determined from the zero-field $\mu^+$SR measurement. However, we cannot make a further statement about the relation between $T_N$ and $T^*$ because it is difficult to perform standard measurements such as heat capacity and magnetization measurements for a tiny crystal. Further experimental and theoretical works are highly desirable to unveil the relation between the two charactersitic temperatures.

**Conclusions**

In summary, we observed a butterfly-shaped MR in triangular-lattice antiferromagnetic $Ag_2CrO_2$ devices. The butterfly-shaped MR can be seen only when the magnetic field is applied along the $c$-axis. This fact indicates a strong uniaxial anisotropy in $Ag_2CrO_2$. The butterfly-shaped MR takes a maximum value of 15% at around the transition temperature, suggesting that spin fluctuations are essential. The result is well explained by the theoretical model based on the 2D magnetic system with the Ising anisotropy. Rich physics is further expected in such a magnetically frustrated system coupled to conducting electrons. In addition, the large MR at small switching fields obtained in the frustrated spin system would be useful for future device applications.

## References

1. Néel, L., Magnetism and Local Molecular Field. *Science* **174**, 985, https://doi.org/10.1126/science.174.4013.985 (1971).

2. Dieny, B., Speriosu, V. S., Gurney, B.A., Parkin, S. S. P., Wilhoit, D. R., Roche, K. P., Metin, S., Peterson, D. T. & Nadimi, S. Spin-valve effect in soft ferromagnetic sandwiches. *J. Magn. Magn. Mater.* **93**, 101, https://doi.org/10.1016/0304-8853(91)90311-W (1991).

3. Maekawa, S., *Concepts in Spin Electronics*, Oxford University Press (2006).

4. Baltz, V., Manchon, A., Tsoi, M., Moriyama, T., Ono, T. & Tserkovnyak, Y. Antiferromagnetic spintronics. *Rev. Mod. Phys.* **90**, 015005, https://doi.org/10.1103/RevModPhys.90.015005 (2018).

5. Jungwirth, T., Sinova, J., Manchon, A., Marti, X., Wunderlich, J. & Felser, C. The multiple directions of antiferromagnetic spintronics. *Nat. Phys.* **14**, 200, https://doi.org/10.1038/s41567-018-0063-6 (2018).

6. Jungfleisch, M. B., Zhang, W. & Hoffmann, A. Perspectives of antiferromagnetic spintronics. *Phys. Lett. A* **382**, 865, https://doi.org/10.1016/j.physleta.2018.01.008 (2018).

7. Wadley, P., Howells, B., Železný, J., Andrews, C., Hills, V., Campion, R. P., Novák, V., Olejník, K., Maccherozzi, F., Dhesi, S. S., Martin, S. Y., Wagner, T., Wunderlich, J., Freimuth, F., Mokrousov, Y., Kuneš, J., Chauhan, J. S., Grzybowski, M. J., Rushforth, A. W., Edmonds, K. W., Gallagher, B. L. & Jungwirth T. Electrical switching of an antiferromagnet. *Science* **351**, 587,

https://doi.org/10.1126/science.aab1031 (2016).

8. Shiino, T., Oh, S.-H., Haney, P. M., Lee, S.-W., Go, G., Park, B.-G. & Lee, K.-J. Antiferromagnetic Domain Wall Motion Driven by Spin-Orbit Torques. *Phys. Rev. Lett.* **117**, 087203, https://doi.org/10.1103/PhysRevLett.117.087203 (2016).

9. Kim, K.-J., Kim, S. K., Hirata, Y., Oh, S.-H., Tono, T., Kim, D.-H., Okuno, T., Ham, W. S., Kim, S., Go, G., Tserkovnyak, Y., Tsukamoto, A., Moriyama, T., Lee, K.-J. & Ono, T. Fast domain wall motion in the vicinity of the angular momentum compensation temperature of ferrimagnets. *Nat. Mater.* **16**, 1187, https://doi.org/10.1038/nmat4990 (2017).

10. Chen, H., Niu, Q. & MacDonald, A. H. Anomalous Hall effect arising from noncollinear antiferromagnetism. *Phys. Rev. Lett.* **112**, 017205, https://doi.org/10.1103/PhysRevLett.112.017205 (2014).

11. Zhang, W., Han, W., Yang, S.-H., Sun, Y., Zhang, Y., Yan, B. & Parkin, S. S. P. Giant facet-dependent spin-orbit torque and spin Hall conductivity in the triangular antiferromagnet $IrMn_3$. *Sci. Adv.* **2**, e1600759, https://doi.org/10.1126/sciadv.1600759 (2016).

12. Kampfrath, T., Sell, A., Klatt, G., Pashkin, A., Mährlein, S., Dekorsy, T., Wolf, M., Fiebig, M., Leitenstorfer A. & Huber R. Coherent terahertz control of antiferromagnetic spin waves. *Nat. Photonics* **5**, 31, https://doi.org/10.1038/nphoton.2010.259 (2011).

13. Khymyn, R., Lisenkov, I., Tiberkevich, V., Ivanov, B. A. & Slavin, A. Antiferromagnetic THz-frequency Josephson-like Oscillator Driven by Spin Current. *Sci.*

*Rep*. **7**, 43705, https://doi.org/10.1038/srep43705 (2017).

14. Nakatsuji, S., Kiyohara, N. & Higo, T. Large anomalous Hall effect in a non-collinear antiferromagnet at room temperature. *Nature* **527**, 212, https://doi.org/10.1038/nature15723 (2015).

15. Kiyohara, N., Tomita, T. & Nakatsuji, S. Giant Anomalous Hall Effect in the Chiral Antiferromagnet Mn3Ge, *Phys. Rev. Applied* **5**, 064009, https://doi.org/10.1103/PhysRevApplied.5.064009 (2016).

16. Nayak, A. K., Fischer, J. E., Sun, Y., Yan, B., Karel, J., Komarek, A. C., Shekhar, C., Kumar, N., Schnelle, W., Kübler, J., Felser C. & Parkin S. S. P. Large anomalous Hall effect driven by a nonvanishing Berry curvature in the noncolinear antiferromagnet Mn3Ge. *Sci. Adv.* **2**, e1501870, https://doi.org/10.1126/sciadv.1501870 (2016).

17. Zhao, K., Hajiri, T., Chen, H., Miki, R., Asano, H. & Gegenwart P. Anomalous Hall effect in the noncollinear antiferromagnetic antiperovskite $Mn_3Ni_{1-x}Cu_xN$. *Phys. Rev. B* **100**, 045109, https://doi.org/10.1103/PhysRevB.100.045109 (2019).

18. Nagaosa, N., Sinova, J., Onoda, S. MacDonald, A. H. & Ong N. P. Anomalous Hall effect. *Rev. Mod. Phys.* **82**, 1539, https://doi.org/10.1103/RevModPhys.82.1539 (2010).

19. Kimata, M., Chen, H., Kondou, K., Sugimoto, S., Muduli, P. K., Ikhlas, M., Omori, Y., Tomita, T., MacDonald, A. H., Nakatsuji, S. & Otani, Y. Magnetic and magnetic inverse spin Hall effects in a non-collinear antiferromagnet. *Nature* **565**, 627, https://doi.org/10.1038/s41586-018-0853-0 (2019).

20. Muduli, P. K., Higo, T., Nishikawa, T., Qu, D., Isshiki, H., Kondou, K., Nishio-Hamane, D., Nakatsuji, S. & Otani, Y. Evaluation of spin diffusion length and spin Hall angle of the antiferromagnetic Weyl semimetal $Mn_3Sn$. *Phys. Rev. B* **99**, 184425, https://doi.org/10.1103/PhysRevB.99.184425 (2019).

21. Yoshida, H., Takayama-Muromachi, E. & Isobe, M. Novel $S = 3/2$ Triangular Antiferromagnet $Ag_2CrO_2$ with Metallic Conductivity. *J. Phys. Soc. Jpn.* **80**, 123703, https://doi.org/10.1143/JPSJ.80.123703 (2011).

22. Matsuda, M., de la Cruz, C., Yoshida, H., Isobe, M. & Fishman R. S. Partially disordered state and spin-lattice coupling in an $S = 3/2$ triangular lattice antiferromagnet $Ag_2CrO_2$. *Phys. Rev. B* **85**, 144407, https://doi.org/10.1103/PhysRevB.85.144407 (2012).

23. Sugiyama, J., Nozaki, H., Miwa, K., Yoshida, H., Isobe, M., Prša, K., Amato, A., Andreica, D. & Månsson M. Partially disordered spin structure in $Ag_2CrO_2$ studied with $\mu^+$SR. *Phys. Rev. B* **88**, 184417, https://doi.org/10.1103/PhysRevB.88.184417 (2013).

24. Kida, T., Okutani, A., Yoshida, H. & Hagiwara, M. Transport Properties of the Metallic Two-dimensional Triangular Antiferromagnet $Ag_2CrO_2$. *Phys. Proc.* **75**, 647, https://doi.org/10.1016/j.phpro.2015.12.083 (2015).

25. Taniguchi, H., Suzuki, S., Arakawa, T., Yoshida, H., Niimi, Y. & Kobayashi, K. Fabrication of thin films of two-dimensional triangular antiferromagnet $Ag_2CrO_2$ and their transport properties. *AIP Adv.* **8**, 025010, https://doi.org/10.1063/1.5016428 (2018).

26. Seki, S., Onose, Y. & Tokura, Y. Spin-Driven Ferroelectricity in Triangular Lattice Antiferromagnets $ACrO_2$ (A = Cu , Ag, Li, or Na). *Phys. Rev. Lett.* **101**, 067204, https://doi.org/10.1103/PhysRevLett.101.067204 (2008).

27. Bass, J. & Pratt Jr., W. P. Current-perpendicular (CPP) magnetoresistance in magnetic metallic multilayers. *J. Mag. Mag. Mater.* **200**, 274, https://doi.org/10.1016/S0304-8853(99)00316-9 (1999).

28. Žutić, I., Fabian, J. & Das Sarma S. Spintronics: Fundamentals and applications. *Rev. Mod. Phys.* **76**, 323 https://doi.org/10.1103/RevModPhys.76.323 (2004).

29. Wegrowe, J.-E., Kelly, D., Franck, A., Gilbert, S. E. & Ansermet, J.-Ph. Magnetoresistance of Ferromagnetic Nanowires. *Phys. Rev. Lett.* **82**, 3681, https://doi.org/10.1103/PhysRevLett.82.3681 (1999).

30. Zhang, D., Ishizuka, H., Lu, N., Wang, Y., Nagaosa, N., Yu, P. & Xue, Q.-K. Anomalous Hall effect and spin fluctuations in ionic liquid gated $SrCoO_3$ thin films. *Phys. Rev. B* **97**, 184433, https://doi.org/10.1103/PhysRevB.97.184433 (2018).

31. Masuda, H., Sakai, H., Tokunaga, M., Yamasaki, Y., Miyake, A., Shiogai, J., Nakamura, S., Awaji, S., Tsukazaki, A., Nakao, H., Murakami, Y., Arima, T., Tokura, Y. & Ishiwata, S. Quantum Hall effect in a bulk antiferromagnet $EuMnBi_2$ with magnetically confined two-dimensional Dirac fermions. *Sci. Adv.* **2**, e1501117, https://doi.org/10.1126/sciadv.1501117 (2016).

32. See Supplementary Information for details on our theoretical model, which includes Refs. 33-37.

33. Lehmann, W. P., Breitling, W. & Weber, R. Raman scattering study of spin dynamics in the quasi-1D Ising antiferromagnets CsCoCl$_3$ and CsCoBr$_3$. *J. Phys. C: Solid State Phys.* **14**, 4655, https://doi.org/10.1088/0022-3719/14/31/014 (1981).

34. Nagler, S. E., Buyers, W. J. L., Armstrong, R. L. & Briat, B. Ising-like spin-1/2 quasi-one-dimensional antiferromagnets: Spin-wave response in CsCoX$_3$ salts. *Phys. Rev. B* **27**, 1784, https://doi.org/10.1103/PhysRevB.27.1784 (1983).

35. Matsubara, F., Inawashiro, S. & Ohhara, H. On the magnetic Raman scattering in CsCoCl$_3$, CsCoBr$_3$ and RbCoCl$_3$. *J. Phys.: Condens. Matter* **3**, 1815, https://doi.org/10.1088/0953-8984/3/12/012 (1991).

36. Todoroki, N. & Miyashita, S. Ordered Phases and Phase Transitions in The Stacked Triangular Antiferromagnet CsCoCl$_3$ and CsCoBr$_3$. *J. Phys. Soc. Jpn.* **73**, 412, https://doi.org/10.1143/JPSJ.73.412 (2004).

37. Koseki, O. & Matsubara, F. Monte Carlo Simulation of Magnetic Ordering of Quasi-One Dimensional Ising Antiferromagnets CsCoBr$_3$ and CsCoCl$_3$. *J. Phys. Soc. Jpn.* **69**, 1202, https://doi.org/10.1143/JPSJ.69.1202 (2000).

**Acknowledgements**

We thank fruitful discussions with S. Maekawa, T. Ziman, B. Gu, M. Hagiwara, T. Kida, Y. Narumi, N. Hanasaki, H. Sakai, H. Murakawa, H. Kawamura, K. Aoyama, and C. Hotta. We also acknowledge the stimulated discussion in the meeting of the Cooperative Research Project of the RIEC, Tohoku University. This work was supported by JSPS KAKENHI (Grant Numbers JP16H05964, JP17K18756, JP26103002, JP19H00656, JP18H04222, JP19K14649, and JP18K03529), the Mazda Foundation, Shimadzu Science Foundation, Yazaki Memorial Foundation for Science and Technology, SCAT Foundation, and the Murata Science Foundation.

**Author contributions**

Y. N. designed the experiments. H. T., S. S., and E. I. fabricated devices and H. T., M. W., M. T., T. I., and T. A. performed the transport measurements and analyzed the data. H. Y. provided polycrystalline $Ag_2CrO_2$ samples. H. I. performed the theoretical calculations. H. T., H. I., K. K., and Y. N. wrote the manuscript. All the authors discussed the results and commented on the manuscript.

**Author contributions**

The authors declare no competing interests.

**Supplementary Information**

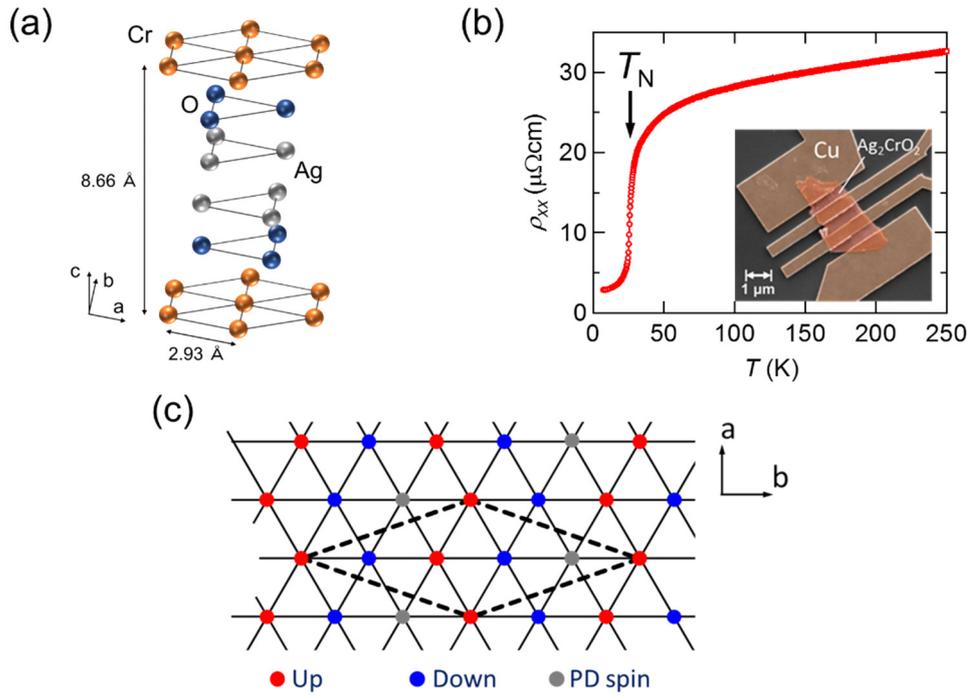

**Figure 1**. **(a)** The crystal structure of $Ag_2CrO_2$. The distance between the two Cr sites in the plane is 2.93 Å. The lattice constant along the *c*-axis is 8.66 Å. **(b)** Temperature dependence of $\rho_{xx}$ of a $Ag_2CrO_2$ thin film. The inset is a scanning electron microscope image of the $Ag_2CrO_2$ device. **(c)** Schematic of the PD state with 5 sublattices. The red, blue, and gray circles represent the up- and down- and PD-spins along the *b*-axis, respectively.

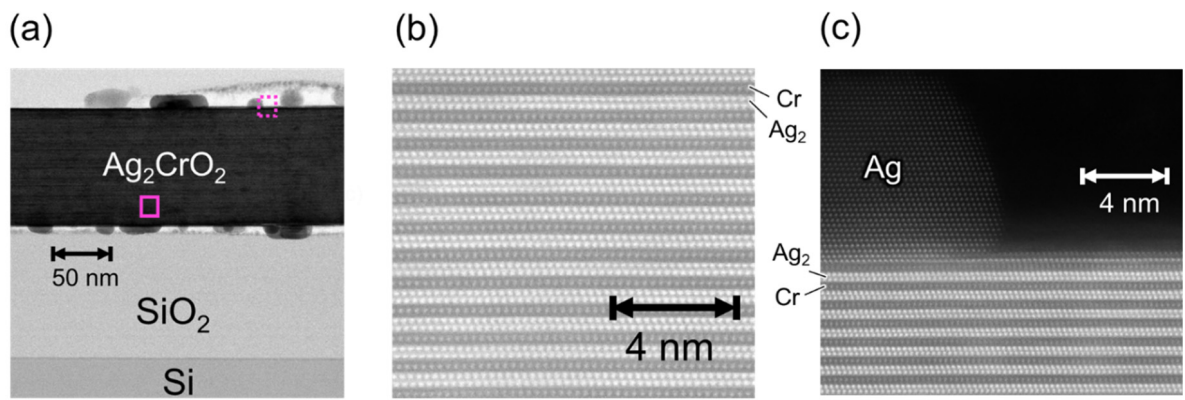

**Figure 2.** STEM images of a typical $Ag_2CrO_2$ device. **(a)** Bright-field STEM image of a wide area. **(b)** High angle annular dark-field STEM image in the area shown with the solid square in **(a)**. The bright and dark spheres correspond to Ag and Cr atoms, respectively. **(c)** High angle annular dark-field STEM image in the area shown with the broken square in **(a)**.

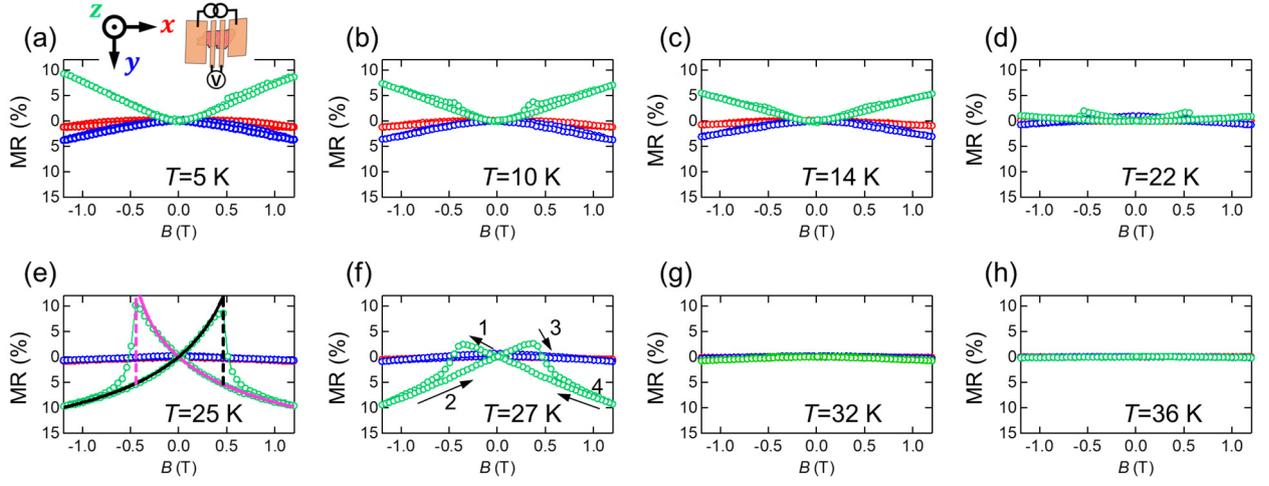

**Figure 3. (a)-(h)** MR curves at several different temperatures ($T$ = 5, 10, 14, 22, 25, 27, 32, and 36 K). The red, blue, and green curves show the MRs when $B$ is applied along the $x$, $y$ and $z$-axes, respectively. The axes are defined as shown in the inset of **(a)**. The solid and broken lines in **(e)** are the best fits with Eq. (2). Because of the uniaxial anisotropy, the MR has a jump at $B \approx \pm 0.5$ T. The arrows and numbers in **(f)** indicate the order of the field sweep direction.

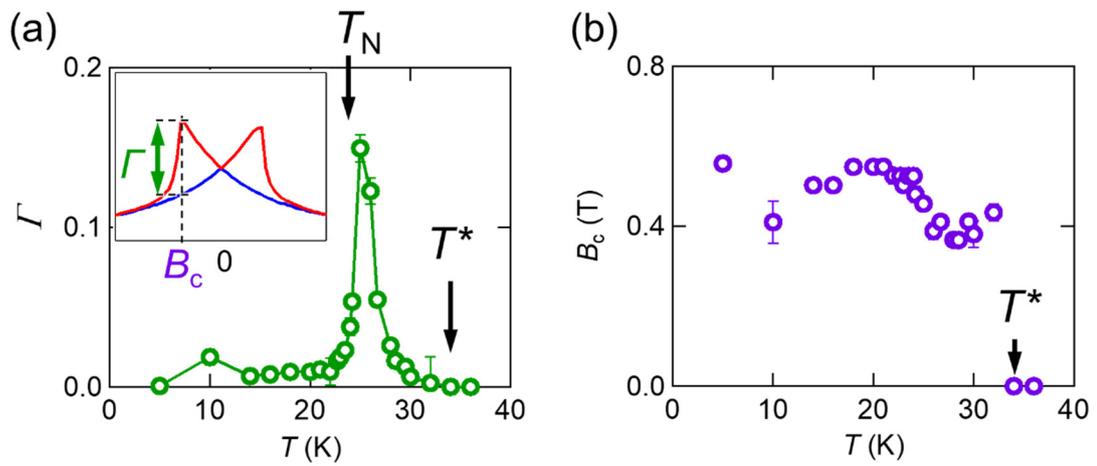

**Figure 4. (a)** The amplitude of the buttery-shaped MR ($\Gamma$) as a function of temperature. The inset shows the definitions of $\Gamma$ and $B_c$. The red and blue curves show $\rho_{xx}^{upper}$ and $\rho_{xx}^{lower}$, respectively. **(b)** Temperature dependence of $B_c$.

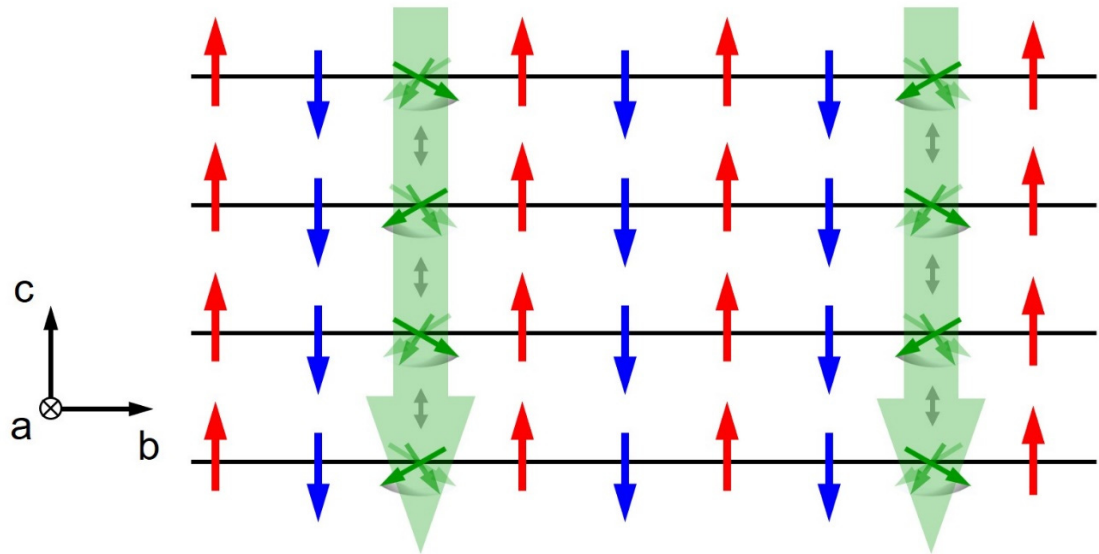

**Figure 5.** Schematic drawing of the spin configuration on the *b-c* plane (see Fig. 1(a)) expected from the experimental result.

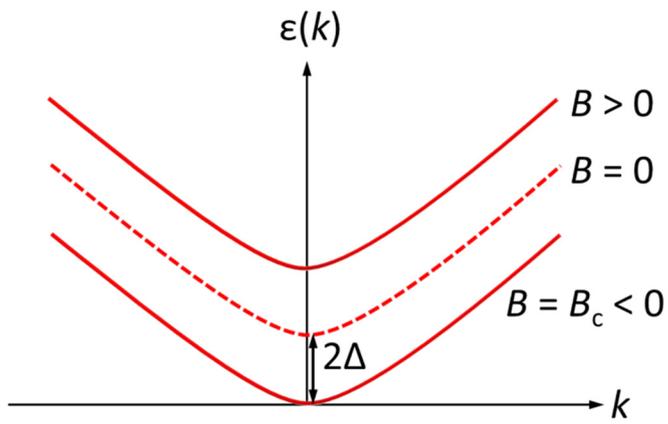

**Figure 6.** Dispersion relation of magnons with the uniaxial anisotropy for the positive magnetization. Because of the anisotropy energy 2Δ, the energy band is shifted and becomes zero when $B = B_c$ (<0).